\documentclass[11pt,twoside]{article}
\usepackage{asp2006}
\usepackage{epsf}
\usepackage{graphicx}
\begin{document}
\title{Composition of the galactic center star cluster}
\author{Rainer M. Buchholz\altaffilmark{1}, Rainer Sch\"odel\altaffilmark{2}, Andreas Eckart\altaffilmark{1,3}}
\altaffiltext{1}{I. Physikalisches Institut, Universit\"at zu K\"oln, Z\"ulpicher Str. 77, 50937 K\"oln, Germany, email: buchholz,eckart@ph1.uni-koeln.de}
\altaffiltext{2}{Instituto de Astrof\'isica de Andaluc\'ia (IAA)-CSIC, Glorieta de la Astronom\'ia s/n, E-18008 Granada, Spain, email: rainer@iaa.es}
\altaffiltext{3}{Max-Planck-Institut f\"ur Radioastronomie, Auf dem H\"ugel 69, 53121 Bonn, Germany}
\begin{abstract}
We present a population analysis of the nuclear stellar cluster of the Milky Way based on adaptive optics narrow band spectral energy distributions. We find strong evidence for the lack of a stellar cusp and a similarity of the late type luminosity function to the bulge KLF.
\end{abstract}
\section{Observations and classification}
We observed the central parsec of the Galactic center using an H band filter and seven intermediate band filters covering the K band with the AO assisted instrument NACO at the ESO VLT on Paranal. This yielded eight point spectral energy distributions (SEDs) for 5914 sources. The spectral feature we used for stellar classification is the CO bandhead absorption ($\lambda > 2.24 \mu$m), which allows the separation of late and early type stars. We determined its presence by fitting the SEDs with an extincted blackbody. Our method produces reliable results for stars as faint as 15.5 mag in the K band (K2III or B2V stars, considering extinction and distance modulus), much deeper than previous studies. We classfied 322 stars as early type candidates and 2955 as late type.
\section{Results}
\begin{figure}
\centering
\includegraphics[scale=0.65]{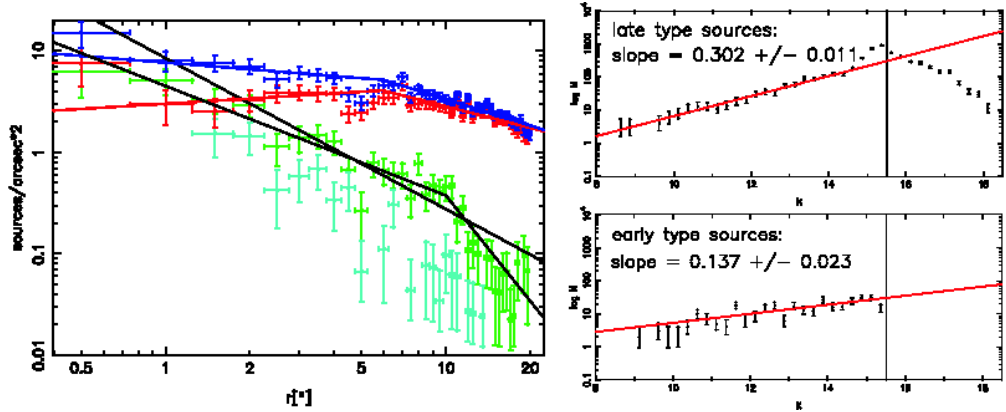}
\caption{\small Left: Projected densities of different classes of sources, green: early type stars, fitted with $\beta = -1.49 \pm 0.12$, resp. $\beta_1 = -1.1 \pm 0.1$ and $\beta_2 = -3.5 \pm 0.6$ for $R_1 < 10'' < R_2$, red: late type stars, fitted with $\beta_1 = 0.17 \pm 0.09$ and $\beta_2 = -0.70 \pm 0.09$ for $R_1 < 6'' < R_2$, blue: all identified stars, light blue: \cite{paumard2006} sources, Right: KLFs for different populations in the GC, with fitted power laws.}
\label{Fig1}
\end{figure}

For the first time, it is now possible to study the early and late type population separately at this depth. Only the latter can be expected to be dynamically relaxed due to its age (Gyrs). Our new results show that the late type population indeed lacks the predicted cusp structure, showing a core instead \citep[see e.g.][]{merritt2009}. The steep decline of the projected early type density towards the outer parts of the cluster \citep[e.g.][]{genzel2003,paumard2006} could be confirmed here as well (for both see Fig.\ref{Fig1}).
The total K band luminosity function of the central parsec is considerably flatter than that of the galactic bulge, with a power law index $\alpha_{GC} = 0.23 \pm 0.02$ \citep{schoedel2007} compared to $\alpha_{bulge} = 0.3$ \citep{figer2004}. The power law slope we fitted to the late type KLF alone agrees very well with that of the bulge, a surprising result since the GC cluster is assumed to be a dynamically separate entity (see Fig.\ref{Fig1}). We can also confirm the much flatter KLF found for the early type stars \citep[e.g.][]{paumard2006} to a much greater depth (K$<$15.5 instead of K$<$13-14).   
\section{Conclusions}
These new results have enhanced our knowledge of the stellar population in the GC considerably, although several points remain unclear. Instead of a cusp, we find a depletion of giants in the very center, as already found by e.g. \cite{haller1996}, for giants with K$<$12, with \cite{do2009,bartko2010} confirming our own results spectroscopically only recently. Several possible causes for this have been suggested, such as collisions/envelope stripping, mass segregation and IMBH/SMBH infall scenarios \citep[see e.g.][]{dale2009}. The fact that the power law slopes of the bulge and the late type KLF agree within errors points to a similar mass function and stellar evolution. How this agrees with both being dynamically separate entities needs to be determined. For our further results and a more detailed treatment of the issues mentioned above, please see \cite{buchholz2009}.

\end{document}